\documentstyle[epsfig,preprint,aps]{revtex}     
\input{epsf}    
    
\newcommand{\vcb}{V_{\rm cb}}

\newcommand{\be}{\begin{eqnarray}}    
\newcommand{\ee}{\end{eqnarray}}    

\newcommand{\g}{\gamma}

\renewcommand{\to}{\rightarrow}    
 
\def\BaBar{{\sl B{\small \sl A}B{\small \sl AR}}} 
 
\def\ifb{{\rm fb^{-1}}} 
\def\fDs{f_{D_s}} 
\def\BR{{\cal B}}

\begin{document}    
    
\draft    
    
\setcounter{page}{0}    
 
\begin{titlepage} 

    
    
\begin{center}    
{\large\bf Test of Factorization Hypothesis \\ 
from Exclusive Non-leptonic $B$-Decays   }    
\end{center}    

    
\begin{center}    
{\sc   C.~ S.~Kim$^{~\mathrm{a}}$\footnote{cskim@mail.yonsei.ac.kr,
~~http://phya.yonsei.ac.kr/\~{}cskim/},
Y.~ Kwon$^{~\mathrm{a}}$\footnote{kwon@phya.yonsei.ac.kr},
Jake~ Lee$^{~\mathrm{a}}$\footnote{jilee@theory.yonsei.ac.kr} and      
W.~ Namgung$^{~\mathrm{b,c}}$\footnote{ngw@dongguk.ac.kr}  } 
           
\begin{small}     
$^{\mathrm{a}}$ {\it Department of Physics and IPAP, Yonsei    
University, Seoul 120-749,  Korea}    
       
$^{\mathrm{b}}$ {\it Department of Physics, Dongguk University, Seoul
100-715,     Korea}   
    
$^{\mathrm{c}}$ {\it School of Physics, Korea Institute for Advanced Study,
Seoul 130-012,     Korea}   
     
\end{small}    
\end{center}    
          
\vspace{-0.5cm}    
   
\setcounter{footnote}{0}    
\begin{abstract}   
We  investigated the possibility of testing factorization
hypothesis in non-leptonic exclusive decays of $B$-meson. 
In particular, we considered the  non-factorizable 
$\overline{B}^0\to D^{(*)+} D_s^{(*)-}$ modes and
$\overline{B}^0\to D^{(*)+}(\pi^-,\rho^-)$ known as well-factorizable modes. 
By taking the ratios  
${\cal B}(\overline{B}^0\to D^{(*)+}D_s^{(*)-})/{\cal B}(\overline{B}^0\to    
D^{(*)+}(\pi^-,\rho^-))$,  
we found that under the present theoretical and experimental uncertainties
there's no evidence for the breakdown of factorization description
to heavy-heavy decays of the $B$ meson.   

\end{abstract}    
    
\end{titlepage}     
    
\newpage    
    
\renewcommand{\thefootnote}{\alph{footnote}}    
    
\section{Introduction}    

\noindent    
Non-leptonic decays of heavy mesons are very important weak processes
for the determination of the Cabibbo-Kobayashi-Maskawa (CKM) matrix
elements \cite{CKM} and the understanding of the CP violation
mechanism.  The analysis of decay processes of $B$-mesons, now
produced in large numbers in the $B$-factories worldwide, will
contribute to the stringent test of the Standard Model (SM) and search
for new physics.  Non-leptonic decays of $B$-mesons, however, are
complicated processes due to inherent hadronic nature and final state
interactions.

A simple formulation of the decay amplitude, so called the naive
factorization scheme \cite{naivefac,BSW}, has been widely used without
full theoretical justification.  And its phenomenological extension,
the generalized factorization scheme, with process-dependent
quantities from penguin effects and non-factorizable contributions has
been also widely used in the literature \cite{genfac,cskim}.  In this latter
scheme, the non-factorizable effects are contained in the effective
color number, $N_c^{\rm eff}$, which is a free parameter
\cite{genfac,chen} of the scheme; the value of $N_c^{\rm eff}$ was
adjusted to $\infty$ for $D$ decays, and to 2 or 5 depending on the
chiral structure of $B$ decays.  However, the $N_c^{\rm
eff}$-dependence of the Wilson coefficient in the effective
Hamiltonian is different for each individual coefficient.  For
example, the coefficients related to color-favored processes, $a_1$,
$a_4$, $a_6$, $a_9$ are stable against the variation of $N_c^{\rm
eff}$, while those related to color-suppressed processes strongly
depend on $N_c^{\rm eff}$ \cite{genfac,cskim,chen}.  This observation
indicates that the non-factorizable contributions in the color-favored
decays are negligible compared with the factorizable contributions.

Recently much progress has been achieved \cite{beneke,pQCD} towards
understanding non-leptonic decay processes by separating out short
distance physics from long distance effects in the well-defined
manner; Beneke {\it et al.} \cite{beneke} proved the validity of
factorization for the $B$-meson decay amplitude in the context of
perturbative QCD formalism.  They showed that when a $B$-meson decays
weakly to a heavy meson and emits a light meson, the decay amplitude
factorizes in the same form as the naive factorization formula, but
with calculable coefficients in the heavy quark limit.  In the case of
$B$-meson decaying to light meson rather than heavy one, according to
their formalizm, a contribution by hard spectator quark is added to
the amplitude, therefore the total amplitude is still factorizable.
However, for $B$-decays to a heavy or light meson with emitting a
heavy meson, their amplitudes are not written as factorized forms,
since the color transparency arguments cannot be applied for such
decays.

Though the factorization of the decay amplitude for a $B$-meson
decaying to heavy-heavy mesons has not been justified, there have been
many calculations using the factorized formula in the literature
\cite{cskim,cheng,rosner}. Within the naive factorization scheme, Luo and
Rosner \cite{rosner} calculated the branching ratios of the $B$-meson
decays, $\overline{B}^0\to D^{(*)+}D_s^{(*)-}$, after extracting the
values of $|\vcb|$ and the slope of the universal Isgur-Wise form
factor $\rho^2$, by comparing the decay rates of $\overline{B}^0\to
D^{(*)+}(\pi^-,\rho^-)$ with a differential distribution of
$\overline{B}^0\to D^{(*)+}l^-\bar{\nu}_l$ measured by the CLEO
Collaboration \cite{recentCLEO}.  They found theoretical predictions
of the naive factorization approach are acceptable within present
experimental errors.
 
Here we test the generalized factorization scheme for the
color-favored $B$-meson decay to heavy-heavy mesons by comparing with
the $B$-decay to heavy-light mesons. The selected decay processes are
$\overline{B}^0\to D^{(*)+} D_s^{(*)-}$ for heavy-heavy and
$\overline{B}^0\to D^{(*)+}(\pi^-,\rho^-)$ for heavy-light, whose
experimental branching ratios are well known.  Compared to the work of
Luo and Rosner, in which the authors used the naive factorization
scheme neglecting penguin effects, we here include penguin effects and
take ratios of the decay rates to reduce the form factor dependence
and cancel the CKM matrix elements.  In the next section, we will give
theoretical descriptions of $\overline{B}^0\to D^{(*)+}(\pi^-,\rho^-)$
and $\overline{B}^0\to D^{(*)+} D_s^{(*)-}$ within generalized
factorization scheme, and present numerical analyses.  Section III
discusses experimental feasibility of our analyses and closes with a
brief summary.

    
\section{Theoretical Description of 
$\overline{B}^0\to D^{(*)+}(\pi^-,\rho^-)$ and $\overline{B}^0\to
D^{(*)+} D_s^{(*)-}$ within Generalized Factorization Approach}    
    
\noindent    
As previously mentioned, there has been general consensus on the
applicability of factorization approach to the color-favored
heavy-light decays $B\to D^{(*)}(\pi,\rho)$, and this has been
recently justified within perturbative QCD formalism \cite{beneke}.
However, questions still remain about the applicability of
factorization to heavy-heavy decay of $B$-meson, such as $B\to
D^{(*)}D_s^{(*)}$.  We here investigate the validity of factorization
hypothesis by taking ratios of branching fractions of {\it presumably}
non-factorizable $\overline{B}^0\to D^{(*)+} D_s^{(*)-}$ modes to
factorizable $\overline{B}^0\to D^{(*)+}(\pi^-,\rho^-)$ modes.

Based on the generalized factorization formalism, the decay amplitudes of our 
interest are  expressed as    
\be    
A(\overline{B}^0\to D^{(*)+}M^-)=    
\frac{G_F}{\sqrt{2}}V_{cb}V_{qq'}^*\tilde{a}(D^{(*)}M)    
            \langle M^-|\bar{q'}\g^\mu (1-\gamma_5) q|0\rangle     
     \langle D^{(*)+} |\bar{c}\g_\mu (1-\gamma_5) b|\overline{B}^0\rangle ,
\label{amp}
\ee    
where $q(q')=u(d)$ for $M=\pi,\rho$ and  $q(q')=c(s)$ for $D_s^{(*)}$. 
The coefficient $\tilde{a}$ includes penguin effects and possible 
non-factorizable contributions in the generalized factorization
scheme.
They are given, neglecting W-exchange diagram and using 
 $V_{tb}V_{ts}^*\cong -V_{cb}V_{cs}^*$, as 
\be  
\tilde{a}(D^{(*)}(\pi,\rho))&=&a_1,\nonumber\\   
\tilde{a}(DD_s)&=&a_1\left(     
          1+\frac{a_4+a_{10}}{a_1}    
 +2\frac{a_6+a_8}{a_1}\frac{m^2_{D_s}}{(m_b-m_c)(m_c+m_s)}\right),\nonumber\\    
\tilde{a}(D^*D_s)&=&a_1\left(     
            1+\frac{a_4+a_{10}}{a_1}    
 -2\frac{a_6+a_8}{a_1}\frac{m^2_{D_s}}{(m_b+m_c)(m_c+m_s)}\right),\nonumber\\    
\tilde{a}(D^{(*)}D_s^*)&=&a_1\left(     
            1+\frac{a_4+a_{10}}{a_1}\right).    
\label{fac_a} 
\ee    
where $a_j$'s represent conventional effective parameters defined as
$a_{2i}=c_{2i}^{\rm eff}+c_{2i-1}^{\rm eff}/N_c^{\rm eff}$ and
$a_{2i-1}=c_{2i-1}^{\rm eff}+c_{2i}^{\rm eff}/N_c^{\rm eff}$.  Using
the numerical values of $a_j$'s of Ref.~ \cite{cheng}, the effective
parameters $\tilde{a}$ defined above are related to $a_1$ by
\be    
|\tilde{a}(B\to DD_s)|&=&0.847a_1,\nonumber\\    
|\tilde{a}(B\to D^*D_s)|&=&1.037a_1,\nonumber\\    
|\tilde{a}(B\to D^{(*)}D_s^*)|&=&0.962a_1,   
\label{a1values}   
\ee    
where the values are obtained by choosing $N_c^{\rm eff}=2$ for
$(V-A)(V-A)$ interactions ($i.e.$ for operators $O_{1,2,3,4,9,10}$)
and $N_c^{\rm eff}=5$ for $(V-A)(V+A)$ interactions ($i.e.$ for
operators $O_{5,6,7,8}$) \cite{cheng}.  We note that the ratios,
$|{\tilde a}/a_1|$, are numerically very stable over different
$N_c^{\rm eff}$ values; for example, the numerical deviations are less
than a few $\%$ for $N_c^{\rm eff}=2,3,5$ and $\infty$.  {}From the
relations in Eq.~(\ref{a1values}) one can see that, at the amplitude
level, the penguin contributions to $\overline{B}^0\to D^{*+}D_s^-$
decay ($\sim 3.7\%$) are much smaller than those for
$\overline{B}^0\to D^+D_s^-$ mode ($\sim 15.3\%$).  In fact the
penguin effects on $\overline{B}^0\to D^+D_s^-$ decay are not small
enough to be simply neglected.  As previously mentioned, the penguin
effects are neglected in the analyses of Ref.~ \cite{rosner}.  We will
show that the inclusion of the penguin effect in $\overline{B}^0\to
D^+D_s^-$ mode improves substantially the theoretical prediction to
the experimental value.  For $\overline{B}^0\to D^{*+}D_s^-$ decay
mode, the penguin contribution can be neglected.  This difference of
penguin contributions between the similar decay modes
$\overline{B}^0\to D^+D_s^-$ and $\overline{B}^0\to D^{*+}D_s^-$ is
due to the different chiral structure of the final states: $B\to D^*$
transitions occur through axial vector currents, while $B\to D$
through vector currents.
 
The ratios    
\be    
{\cal R}_{D_s^{(*)}/(\pi,\rho)}&\equiv&     
    \frac{{\cal B}(\overline{B}^0\to D^+ D_s^{(*)-})}    
{{\cal B}(\overline{B}^0\to D^+(\pi^-,\rho^-) )}\\    
{\rm and}~~~~\tilde{\cal R}_{D_s^{(*)}/(\pi,\rho)}&\equiv&     
    \frac{{\cal B}(\overline{B}^0\to D^{*+} D_s^{(*)-})}    
{{\cal B}(\overline{B}^0\to D^{*+}(\pi^-,\rho^-) )}   
\ee    
are given as    
\be    
{\cal R}_{D_s/\pi}=    
\left|\frac{\tilde{a}(DD_s)}{\tilde{a}(D\pi)}\right|^2  
           \left(\frac{f_{D_s}}{f_\pi}\right)^2   
           \left(\frac{p_c^{DD_s}}{p_c^{D\pi}}\right)    
           \left(\frac{F_0^{BD}(m_{D_s}^2)}{F_0^{BD}(m_{\pi}^2)}\right)^2,    
\label{Rpi}    
\ee    
\be    
{\cal R}_{D_s^*/\rho}=    
\left|\frac{\tilde{a}(DD_s^*)}{\tilde{a}(D\rho)}\right|^2  
           \left(\frac{f_{D_s^*}}{f_\rho}\right)^2   
           \left(\frac{p_c^{DD_s^*}}{p_c^{D\rho}}\right)^3   
           \left(\frac{F_1^{BD}(m_{D_s^*}^2)}{F_1^{BD}(m_{\rho}^2)}\right)^2,    
\label{Rrho}    
\ee    
\be    
\tilde{\cal R}_{D_s/\pi}=    
\left|\frac{\tilde{a}(D^*D_s)}{\tilde{a}(D^*\pi)}\right|^2  
           \left(\frac{f_{D_s}}{f_\pi}\right)^2   
           \left(\frac{p_c^{D^*D_s}}{p_c^{D^*\pi}}\right)^3    
           \left(\frac{A_0^{BD^*}(m_{D_s}^2)}{A_0^{BD^*}(m_{\pi}^2)}\right)^2,    
\label{Rtilpi}    
\ee    
\be    
\tilde{\cal R}_{D_s^*/\rho}=    
\left|\frac{\tilde{a}(D^*D_s^*)}{\tilde{a}(D^*\rho)}\right|^2  
           \left(\frac{f_{D_s^*}}{f_\rho}\right)^2   
           \left(\frac{p_c^{D^*D_s^*}}{p_c^{D^*\rho}}\right)  
           \left(\frac{m_{D_s^*}}{m_{\rho}}\right)^2  
           \left(\frac{A_1^{BD^*}(m_{D_s^*}^2)}{A_1^{BD^*}(m_{\rho}^2)}\right)^2 
           \left(\frac{H(m_{D_s^*}^2)}{H(m_{\rho}^2)}\right),    
\label{Rtilrho}    
\ee    
where $p_c^{XY}$ is the c.m. momentum of the decay particles 
and we used $|V_{cs}^*/V_{ud}|=1$.    
Here the form factors have the following parameterization   \cite{BSW}:    
\be    
\langle P^\prime (p^\prime) |V_\mu| P(p)\rangle &=&     
    \left(p_\mu+p^\prime_\mu-\frac{m_P^2-m_{P^\prime}^2}{q^2}q_\mu\right)F_1(q^2)    
             +\frac{m_P^2-m_{P^\prime}^2}{q^2}q_\mu F_0(q^2),\nonumber\\    
\langle V(p^\prime,\epsilon) |V_\mu| P(p)\rangle &=&    
       \frac{2}{m_P+m_V}\epsilon_{\mu\nu\alpha\beta}    
           \epsilon^{*\nu}p^\alpha p^{\prime\beta}V(q^2),\nonumber\\    
\langle V(p^\prime,\epsilon) |A_\mu| P(p)\rangle &=&    
       i\left[(m_P+m_V)\epsilon_\mu A_1(q^2)-\frac{\epsilon\cdot p}{m_P+m_V}    
              (p+p^\prime)_\mu A_2(q^2)\right.\nonumber\\    
    &&\left.-2m_V\frac{\epsilon\cdot p}{q^2}q_\mu [A_3(q^2)-A_0(q^2)]\right],    
\ee    
where $q=p-p^\prime$, $F_1(0)=F_0(0)$, $A_3(0)=A_0(0)$,    
$$A_3(q^2)=\frac{m_P+m_V}{2m_V}A_1(q^2)-\frac{m_P-m_V}{2m_V}A_2(q^2),$$    
and $P$, $V$ denote the pseudoscalar and vector mesons, respectively. 
For $B\to V_1V_2$ decay (see Eq.~(\ref{Rtilrho})),  three form factors 
$A_1(q^2)$, $A_2(q^2)$, and $V(q^2)$ contribute. 
Here we factored out the dominant one $A_1(q^2)$ and the other two are 
put in the function $H(q^2)$ defined as 
\be 
H(q^2)=(a-bx)^2+2(1+c^2y^2), 
\ee 
with 
\be 
&&a=\frac{m_B^2-m_1^2-m_2^2}{2m_1m_2},\quad 
  b=\frac{2m_B^2p_c^2}{m_1m_2(m_B+m_1)^2},\quad  
  c=\frac{2m_Bp_c}{(m_B+m_1)^2},\\ 
&&x=\frac{A_2^{BV_1}(q^2)}{A_1^{BV_1}(q^2)},\quad\quad\quad\quad
  y=\frac{V^{BV_1}(q^2)}{A_1^{BV_1}(q^2)} ,
\ee 
where $m_1~(m_2)$ is the mass of the vector meson $V_1~(V_2)$.  Using
the above ratios Eqs.(\ref{Rpi})-(\ref{Rtilrho}), one can, in
principle, test the validity of factorization without having
dependence on CKM matrix elements.  However, the analysis depends
strongly on nonperturbative hadronic factors such as decay constants
and form factors.  $B\to D^{(*)}$ transition form factors are rather
well-constrained and the uncertainty in their ratios would be rather
moderate.  In the following numerical analysis, we consider three
models for the form factors of $B\to D^{(*)}$ transitions: the
Bauer-Stech-Wirbel (BSW) model \cite{BSW}, Melikhov/Stech \cite{MS},
and relativistic light-front (LF) quark model \cite{LF}.

Another uncertainty comes from decay constants, especially  
$f_{D_s^{(*)}}$, which has presently large uncertainty. 
Particle Data Group report \cite{PDG} 
gives two distinct values depending on its decay modes: 
\be 
f_{D_s^+}&=&194\pm 35 \pm 20\pm 14\;{\rm MeV}\;\;~~~ 
             {\rm from}\; D_s\to \mu \nu_\mu,\\ 
f_{D_s^+}&=&309\pm 58 \pm 33\pm 38\;{\rm MeV}\;\;~~~
             {\rm from}\; D_s\to \tau \nu_\tau. 
\ee 
Recently, a rather interesting value appeared in Ref.~\cite{beatrice}: 
\be 
f_{D_s^+}&=&323\pm 44 \pm 12\pm 34\;{\rm MeV}\;\; 
             {\rm from}\; D_s\to \mu \nu_\mu, 
\ee 
which is obtained by measuring the branching fraction of $D_s\to \mu \nu_\mu$ 
relative to the branching fraction $D_s\to \varphi\pi\to K^+K^-\pi$. 
We use the statistical average of the above three, 
\be 
f_{D_s}&=&252\pm 31 \;{\rm MeV}. 
\ee 
 
Then we get the theoretical predictions    
\be    
{\cal R}_{D_s/\pi} 
           &=&[3.38\pm 0.841] 
           \left|\frac{\tilde{a}(DD_s)}{\tilde{a}(D\pi)}\right|^2  
           \left(\frac{f_{D_s}}{0.252}\right)^2 
            \left(\frac{F_0^{BD}(m_{D_s}^2)}{0.74}\right)^2     
           \left(\frac{0.686}{F_0^{BD}(m_{\pi}^2)}\right)^2,\nonumber\\ 
{\cal R}_{D_s^*/\rho} 
           &=&[0.92\pm 0.236]    
            \left|\frac{\tilde{a}(DD_s^*)}{\tilde{a}(D\rho)}\right|^2  
           \left(\frac{f_{D_s^*}}{0.252}\right)^2   
           \left(\frac{F_1^{BD}(m_{D_s^*}^2)}{0.817}\right)^2 
           \left(\frac{0.701}{F_1^{BD}(m_{\rho}^2)}\right)^2,\nonumber\\       
\tilde{\cal R}_{D_s/\pi} 
           &=&[2.17\pm 0.654]   
            \left|\frac{\tilde{a}(D^*D_s)}{\tilde{a}(D^*\pi)}\right|^2  
           \left(\frac{f_{D_s}}{0.252}\right)^2   
            \left(\frac{A_0^{BD^*}(m_{D_s}^2)}{0.793}\right)^2   
           \left(\frac{0.699}{A_0^{BD^*}(m_{\pi}^2)}\right)^2,\nonumber\\
\tilde{\cal R}_{D_s^*/\rho} 
           &=&[2.15\pm 0.545]
           \left|\frac{\tilde{a}(D^*D_s^*)}{\tilde{a}(D^*\rho)}\right|^2  
           \left(\frac{f_{D_s^*}}{0.252}\right)^2   
            \left(\frac{A_1^{BD^*}(m_{D_s^*}^2)}{0.730}\right)^2 
           \left(\frac{0.673}{A_1^{BD^*}(m_{\rho}^2)}\right)^2, 
\label{Rnaive} 
\ee 
where the quoted errors are based on our estimates of uncertainties in
the form-factor model-dependence and in the decay constants
$f_{D_s^{(*)}}$.  Here we assumed $f_{D_s^*}=f_{D_s}$ for simplicity
and used $f_\pi=131\;{\rm MeV}$ and $f_\rho=209\;{\rm MeV}$
\cite{beneke}.  As the ratios of $\tilde{a}$'s are factored out, the
numerical predictions of Eqs.~(\ref{Rnaive}) correspond to those in
the naive factorization approximation.  As is shown, the main
uncertainty comes from our ignorance on the decay constant
$f_{D_s^{(*)}}$.  Within the generalized factorization scheme and by
including penguin effects, the central values of the ratios are
shifted to
\be    
{\cal R}_{D_s/\pi}^{\rm GF}&=&2.43\pm 0.61         ,\nonumber\\  
{\cal R}_{D_s^*/\rho}^{\rm GF}&=&0.85\pm 0.22      ,\nonumber\\    
\tilde{\cal R}_{D_s/\pi}^{\rm GF}&=&2.33\pm 0.70   ,\nonumber\\   
\tilde{\cal R}_{D_s^*/\rho}^{\rm GF}&=&2.00\pm 0.50,
\label{Rgf} 
\ee 
where we used the explicit numerical values for $\tilde{a}$ of
Eq.~(\ref{a1values}).  Considering the current experimental branching
ratios for each decay mode \cite{rosner,PDG}, one gets the following
ratios:
\be 
{\cal R}^{\rm exp}_{D_s/\pi}&=&2.67\pm 1.061,\nonumber\\ 
{\cal R}^{\rm exp}_{D_s^*/\rho}&=&1.27\pm 0.671,\nonumber\\ 
\tilde{\cal R}^{\rm exp}_{D_s/\pi}&=&3.58\pm 1.138,\nonumber\\ 
\tilde{\cal R}^{\rm exp}_{D_s^*/\rho}&=&2.16\pm 0.817. 
\label{Rexp} 
\ee 
Comparing the ratios (\ref{Rnaive}), (\ref{Rgf}) and (\ref{Rexp}), all
the theoretical predictions are well within the present experimental
constraints. We note that the inclusion of the penquin effects for
$\overline{B}^0\to D_s^-D^+$, which add a sizable contribution,
improves the central value so that it is much closer to the
experimental value.  Although presently the experimental errors are
too large to say anything definite, our analysis indicates that
factorization hypothesis is still a good method for describing the
$B$-meson decaying to heavy-heavy mesons. Furthermore, one could even
consider a possibility that the factorization may not be a consequence
of only perturbative QCD, in contrast to the arguments of
Ref.~\cite{beneke}.  Similar arguments are given in
Ref.~\cite{ligeti}, in which the authors considered $B\to D^{(*)}X$
decays and expected non-factorization effects would grow with the
invariant mass $m_X^2$ of the multi hadronic state X if the
factorization is a consequence of perturbative QCD, but they found no
such dependence on $m_X^2$.

\section{Discussions on experimental feasibility and Summary}

\noindent    
The comparison of (\ref{Rgf}) and (\ref{Rexp}) will give a test of
generalized factorization model that we considered in this paper.  As
it stands now, the two sets of values are consistent well within
uncertainties.  On the theoretical side, the biggest uncertainty is in
the determination of meson decay constants $f_{D_s^{(*)}}$, while on
the experimental side the statistical errors of ${\cal B}(B\rightarrow
D^{(*)} D_s^{(*)})$ give the largest uncertainty.  Therefore, we need
to improve the precision of such experimental measurements, for the
method described in this paper to have any significance.
 
As of this writing, the combined data sample of {\BaBar} and Belle
experiments is more than $30~\ifb$ \cite{recentB}.  The peak
instantaneous luminosity of each experiment is over $3\times
10^{33}{\rm cm}^{-2}{\rm s}^{-1}$, which corresponds to more than
$1~\ifb$/week for each.  At this rate, we expect to have more than
$100~\ifb$ of data accumulated within a year. We will take this as our
basis for considering experimental feasibility.
 
Currently, the most precise measurement of $f_{D_s}$ is obtained by
the CLEO collaboration~\cite{fDsChada} in $D_s \to \mu \nu$ decays.
Adding the errors in quadrature, they obtained $\fDs = 280 \pm 48$;
17~\% total uncertainty (7~\% statistical) in $\fDs$, corresponding to
34~\% error in our calculation of the ratios.  With $100~\ifb$ data
sample from {\BaBar} and Belle, which is more than 20 times that of
the existing result~\cite{fDsChada}, the statistical error will be
reduced to $1/\sqrt{20}$ of Ref.~\cite{fDsChada}.  The systematic
errors may not go down as fast, but better understanding of every
other aspect of the analysis will help reduce the systematic
uncertainties.  Assuming that the systematic error can be reduced to
1/3 of Ref.~\cite{fDsChada}, the $\fDs$ value will be determined to
5~\% accuracy, hence resulting in 10~\% error in our ratio.  As for
the form factor errors in the theoretical calculations, we hope that
in the near future precision measurements in heavy-flavor physics
processes from $B$ or charm factories should help test and confirm the
reliability of lattice QCD technique greatly.  Then we may have much
improved form factor errors.

In the experimental measurements of branching ratios, $B \to
D^{(*)}\pi$ modes are measured with much better precision than $B \to
D^{(*)}\rho$ modes.  Similarly, $D^{(*)} D_s$ modes are determined
with significantly higher precision than $D^{(*)} D_s^*$ modes.
Therefore, we expect that $R_{D_s/\pi}$ and $\tilde{R}_{D_s/\pi}$ will
be determined with higher precision than other ratios.  The most
recent and precise measurement of $\BR (B\to D^{*+} \pi^-) = (2.81 \pm
0.24 \pm 0.05) \times 10^{-3}$ is obtained by the CLEO collaboration
with $3.1~\ifb$ of data sample.  With a data sample of $100~\ifb$, we
can measure this branching ratio with a precision at a few \% level.
On the other hand, the most precise measurement of $\BR(B \to D^{*+}
D_s^-)$ also made by CLEO with $2.0~\ifb$ of data, is $(0.90 \pm 0.27
\pm 0.22) \times 10^{-2}$.  Again, with $100~\ifb$ of data, the
statistical error will be reduced to $\sim 1/7$ of its current value.
If experimental systematic error can be reduced to $\sim 40~\%$ of its
current value, $\BR (B \to D^{*+} D_s^-)$ will be determined to 10~\%
level.  Therefore, it is likely that the experimental value of
$\tilde{R}_{D_s/\pi}$ can be determined at the level of $10~\%$
precision.  Similar case can be made with $R_{D_s/\pi}$.
 
Comparing (\ref{Rgf}) and (\ref{Rexp}),  we note that the experimental 
and  theoretical  values  of  $\tilde{R}_{D_s/\pi}$ show  the  biggest 
difference if we  accept their central values.  We  also note that if 
both values can be determined within $10~\%$ accuracy and if we assume 
that their central  values stand as they are, then we  will be able to 
see $3~\sigma$ difference in $\tilde{R}_{D_s/\pi}$.  In conclusion, we 
will  have a good  opportunity to  test the  generalized factorization 
scheme as discussed in this paper once we have $\sim 100~\ifb$ of data 
from the $B$-factory experiments. 
 
To summarize, we have investigated the possibility of testing
factorization hypothesis from non-leptonic exclusive decays of $B$
meson into two meson final states.  In particular, we considered the
{\it presumably} non-factorizable $\overline{B}^0\to D^{(*)+}
D_s^{(*)-}$ modes and $\overline{B}^0\to D^{(*)+}(\pi^-,\rho^-)$ known
as well-factorizable modes.  By taking the ratios ${\cal
B}(\overline{B}^0\to D^{(*)+}D_s^{(*)-})/{\cal B}(\overline{B}^0\to
D^{(*)+}(\pi^-,\rho^-))$, the dependence on CKM matrix elements
vanishes and some model-dependence on hadronic form-factors is
reduced.  We found that under the present theoretical and experimental
uncertainties there's no evidence for breakdown of factorization
description to heavy-heavy decays of the $B$-meson.

\section*{Acknowledgments}    
    
\noindent    
We thank G. Cvetic for careful reading of the manuscript and his
valuable comments.  The work of C.S.K. was supported in part by BK21
Program, CHEP-SRC Program and Grant No. 2000-1-11100-003-1 of the
KOSEF, and in part by the KRF Grant, Project No. 2000-015-DP0077.  The
work of Y.K. was supported by 2000 Yonsei University Faculty Research
Fund.  The work of J.L. was supported by the KRF Grant, Project No.
2000-015-DP0077.  The work of W.N. was supported by the Basic Science
Research Program of Dongguk University.


\begin{thebibliography}{99}     
     
\bibitem{CKM} N.~Cabibbo, Phys.\ Rev.\ Lett. {\bf 10}, 531 (1963);
 M.~Kobayashi and T.~Maskawa,      
Prog.\ Theor.\ Phys.\ {\bf 49}, 652 (1973).   
 
\bibitem{naivefac} J.~Schwinger, Phys.\ Rev.\ Lett. {\bf 12}, 630 (1964); 
  R.P.~Feynman, {\it in} Symmetries in Particle Physics, ed. A.~Zichichi 
  (Academic Press, New York, 1965), p.~167; 
  O.~Haan and B.~Stech, Nucl.\ Phys.\ B {\bf 20}, 448 (1970); 
  D.~Fakirov and B.~Stech, Nucl.\ Phys. B {\bf 133}, 315 (1978); 
  N.~Cabibbo and L.~Maiani, Phys.\ Lett. B {\bf 73}, 418 (1978); 
   {\bf 76}, 633 (1978); 
  L.L.~Chau, Phys.\ Rep. {\bf 95}, 1 (1983). 
 
\bibitem{BSW} M.~Wirbel, B.~Stech and M.~Bauer,      
Z.\ Phys.\ C {\bf 29}, 637 (1985);     
              M.~Bauer, B.~Stech and M.~Wirbel,      
Z.\ Phys.\ C {\bf 34}, 103 (1987).     
 
\bibitem{genfac} H.-Y.~Cheng, Phys.\ Lett.\ B {\bf 335}, 428 (1994); 
 {\bf 395}, 345 (1997); 
H.-Y.~Cheng and B.~Tseng, Phys.\ Rev. D {\bf 58}, 094005 (1998); 
J.M.~Soares, Phys.\ Rev. D {\bf 51}, 3518 (1995); 
M.~Neubert and B.~Stech, {\it in} Heavy Flavours II, eds. A.J.~Buras and 
M.~Lindner (World Scientific, Singapore), p.~294; 
A.~Ali and C.~Greub, Phys.\ Rev. D {\bf 57}, 2996 (1998).

\bibitem{cskim} C.S. Kim, Y. Kwon, Jake Lee and W. Namgung, Phys. Rev. D {\bf 63}, 094506 (2001). 
 
\bibitem{chen}  Y.-H.~Chen, H.-Y.~Cheng, and B.~Tseng,      
Phys.\ Rev.\  D {\bf 59}, 074003 (1999);  
 Y.-H.~Chen, H.-Y.~Cheng, B.~Tseng, and K.-C.~Yang,      
Phys.\ Rev.\  D {\bf 60}, 094014 (1999).

\bibitem{beneke} M.~Beneke, G.~Buchalla, M.~Neubert and C.T.~Sachrajda, 
Phys.\ Rev.\ Lett. {\bf 83}, 1914 (1999); Nucl.\ Phys. B {\bf 591}, 313 (2000). 
 
\bibitem{pQCD} A.~Szczepaniak, E.M.~Henley, and S.J.~Brodsky, 
Phys.\ Lett.\ B {\bf 243}, 287 (1990); 
Y.Y.~Keum, H.N.~Li, and A.I.~Sanda, hep-ph/0004004; Phys.\ Rev.\ D {\bf 63}, 
054008 (2001); C.D.~L\"u, K.~Ukai, and M.Z.~Yang, Phys.\ Rev.\ D {\bf 63}, 
074009 (2001); C.D.~L\"u, M.Z.~Yang, hep-ph/0011238. 
 
 
\bibitem{cheng} H.-Y.~Cheng and K.-C.~Yang,      
Phys.\ Rev.\  D {\bf 59}, 092004 (1999).     
 
\bibitem{rosner} Z.~Luo and J.L.~Rosner, hep-ph/0101089. 
 
\bibitem{recentCLEO} CLEO Collaboration, J.P.~Alexander {\it et al.},  
CLEO-CONF 00-3, presented at XXX International Conference on High Energy Physics, 
Osaka, Japan, July 27 - August 2, 2000. 
 
\bibitem{MS} D.~Melikhov and B.~Stech, Phys.\ Rev. D {\bf 62}, 014006 (2000).     
 
\bibitem{LF} H.Y.~Cheng, C.Y.~Cheung, and C.W.~Hwang,      
Phys.\ Rev.\ D {\bf 55}, 1559 (1997).    
 
\bibitem{PDG} Particle Data Group, C. Caso {\it et al.},     
        Eur. Phys. J. C {\bf 3}, 1 (1998).   
 
\bibitem{beatrice} BEATRICE Collaboration, Y.~Alexandrov {\it et al.},  
Phys.\ Lett. B {\bf 478}, 31 (2000). 
 
\bibitem{ligeti} Z.~Ligeti, M.~Luke and M.B.~Wise, 
hep-ph/0103020. 
 
\bibitem{recentB} 
Belle Collaboration, A. Abashian {\it et al.},
Phys. Rev. Lett. {\bf 86}, 2509 (2001);
{\BaBar } Collaboration, B. Aubert {\it et al.},
Phys. Rev. Lett. {\bf 86}, 2515 (2001).
    
\bibitem{fDsChada} CLEO Collaboration, M. Chada {\it et al.}, 
Phys. Rev. D {\bf 58}, 032002 (1998).
     
\end{thebibliography}
\end{document}